# Broadband, Lensless and Optomechanically Stabilised Coupling into Microfluidic Hollow-Core Photonic Crystal Fiber Using Glass Nanospike

Richard Zeltner[a,b], Shangran Xie[a], Riccardo Pennetta[a] and Philip St.J. Russell[a,b]

[a]Max Planck Institute for the Science of Light, Staudtst.2, 91058 Erlangen
[b]Cluster of Excellence Engineering of Advanced Materials, University of Erlangen-Nuremberg, Staudtst.2, 91058 Erlangen

**ABSTRACT:** We report a novel technique for launching broadband laser light into liquid-filled hollow-core photonic crystal fiber (HC-PCF). It uniquely offers self-alignment and self-stabilization via optomechanical trapping of a fused silica nanospike, fabricated by thermally tapering and chemically etching a single mode fiber into a tip diameter of 350 nm. We show that a trapping laser, delivering ~300 mW at 1064 nm, can be used to optically align and stably maintain the nanospike at the core center. Once this is done, a broadband supercontinuum beam (~575 to 1064 nm) can be efficiently and close to achromatically launched in the HC-PCF. The system is robust against liquid-flow in either direction inside the HC-PCF and the Fresnel back–reflections are reduced to negligible levels compared to free-space launching or butt-coupling. The results are of potential relevance for any application where the efficient delivery of broadband light into liquid-core waveguides is desired.

Optofluidic waveguides[1] permit efficient interactions between light and liquid samples, and are of great interest in a great variety of applications in the life sciences. Although solid-core waveguides can be efficiently integrated into microfluidic systems[2–5], the overlap between the guided optical mode and the specimen is typically very small[1]. In contrast, waveguides with liquid-filled cores, such as anti-resonant reflecting optical waveguides (ARROWs)[6–8] or hollow-core photonic crystal fiber (HC-PCF)[9–11], provide close to 100% overlap between the guided modes and the specimen, enabling extremely efficient light-matter interactions over long path-lengths. As a consequence they are a natural choice for optofluidic experiments, with applications in photochemistry, spectroscopy and sensing[11–13] as well as optical particle manipulation[14–19]. For liquid-core waveguides, light delivery is typically achieved via free-space coupling using lenses, or by butt-coupling to solid-core waveguides such as step-index fibers or ridge-waveguides[13,20]. Free-space coupling relies on bulky optics and translation stages, requires careful alignment and suffers intrinsically from chromatic aberrations, especially when coupling of a broad spectrum is desired. This becomes in particular critical for smaller core diameters. Moreover, in certain experimental configurations, the coupling may suffer from instabilities or evaporation at the liquid-air surface close to the end-face of the waveguide. For butt-coupling, the liquid-core waveguide can be spliced to the fiber, or both waveguides can be written into the same optofluidic chip[20,21]. Splicing of liquid-core waveguides to solid core waveguides makes readjustment difficult and the solid joint between the waveguides is hardly compatible with microfluidic circulation. More importantly, the coupling is multimode, its efficiency being limited both by the mode field mismatch and, in the case of liquid-filled HC-PCF, by unavoidable distortion and even collapse of the microstructure during the splicing process. Finally, Fresnel reflection occurs over a broad spectral range for both free-space and butt-coupling even when anti-reflection-coatings are used to suppress it at specific wavelengths.

Recently a new technique for launching laser light into air-filled HC-PCF was reported, based on a silica "nanospike" with the unique features of self-alignment and a strongly suppressed Fresnel back-reflection[22]. The nanospike was fabricated by thermally tapering a step-index single mode fiber (SMF) and etching its tip in hydrofluoric acid vapor. The core mode of the SMF converts into the fundamental mode of the tapered glass-air waveguide, which then spreads out into the surrounding air as it travels along the nanospike, adiabatically evolving into the fundamental mode of the surrounding HC-PCF. Optomechanical back-action between nanospike and hollow core produces a strong restoring force on the nanospike, trapping it at core center.

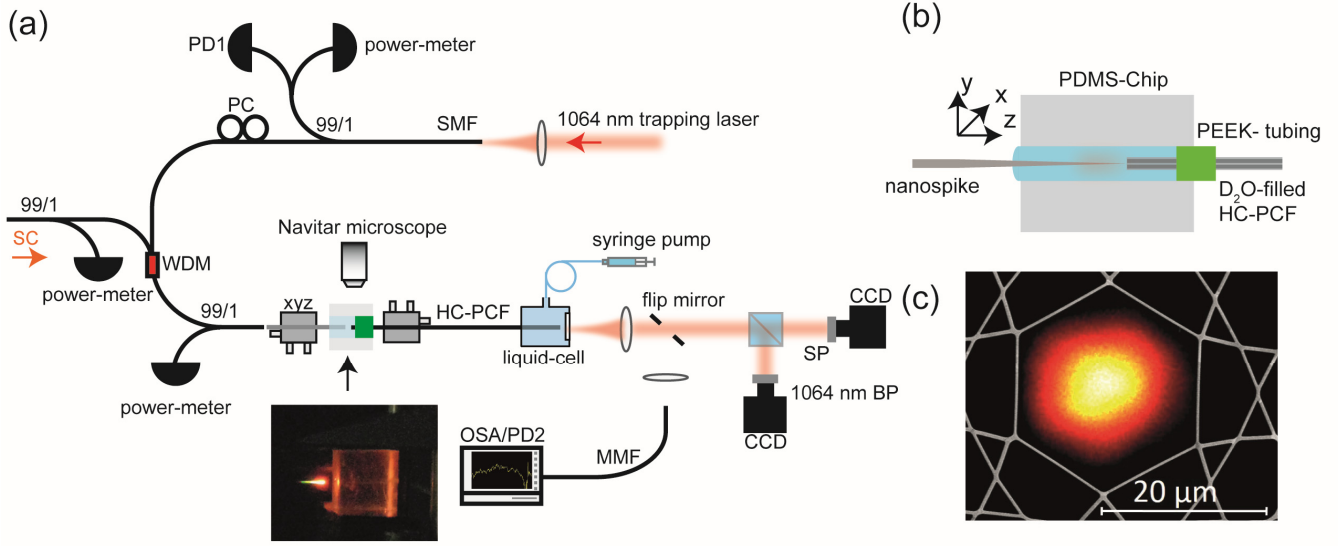

Fig 1. (a) Experimental set-up. PD, photodiode; PC, polarization controller; WDM, wavelength division multiplexer; SC, supercontinuum; SP, short pass filter; BP, band pass filter; OSA, optical spectrum analyzer; MMF, multi-mode fiber. (b) Sketch of the microfluidic channel with the nanospike and the HC-PCF inside it (not to scale). The HC-PCF is moved along the z-axis towards until it envelopes the nanospike. Any angular misalignment between the axis of the nanospike and the fiber core can be corrected by adjustment of the transverse nanospike position (c) Optical near-field image of the excited mode at 1064 nm when the nanospike is at core center, superimposed on a scanning electron micrograph of the fiber core.

Here we report the first successful implementation of this technique in liquid-filled HC-PCF, showing that by use of a relatively strong trapping beam at a single wavelength it is possible to achieve efficient, close to achromatic, coupling from 575 nm to 1064 nm, limited by the bandwidth of the optical components and the HC-PCF. The system is ideal for carrying out optofluidic spectroscopy using HC-PCF. Lensless optical self-alignment reduces the complexity of manual alignment and makes the system robust against external perturbations. The system is also found to be robust against liquid counter-flow through the HC-PCF.

**RESULTS AND DISCUSSION**

The experimental set-up is sketched in Fig. 1a. The central component is a straight microfluidic channel made from polydimethylsiloxane (PDMS) with an internal diameter of ~700 μm and a length of ~1 cm. One end of a ~23 cm long kagomé-style HC-PCF, core diameter 21.5 μm, is mounted on a xyz translation stage and inserted into the microfluidic channel. The other end of the HC-PCF is clamped inside a liquid-cell that is connected to a syringe pump, allowing $D_2O$ to be introduced into both the microfluidic channel and the HC-PCF.

For optimal self-alignment and launch efficiency, the tip diameter must be chosen so that the optical mode diameter approximately matches that of the HC-PCF mode. Using the analytical approximation of the mode field radius of the tip[23] and the HC-PCF[24] this leads to the condition:

$$\frac{a}{R}\left(\alpha_1 + \frac{\alpha_2}{V^{1.5}} + \frac{\alpha_3}{V^6}\right) = \alpha_4 \quad (1)$$

where $\alpha_1$ = 0.65, $\alpha_2$ = 1.62, $\alpha_3$ = 2.88 and $\alpha_4$ = 0.69, $a$ and $R$ are respectively the final tip radius and core radius of the HC-PCF, and $V = 2\pi a \sqrt{n_G^2 - n_L^2}/\lambda$, $\lambda$ being the vacuum wavelength. The smaller index step between glass ($n_G$) and liquid ($n_L$) means that the optimal nanospike tip diameter is greater than in an air-filled PCF with the same core diameter, working out to be ~350 nm in this case. The nanospike was inserted into the other end of the PDMS-chip and, using a second xyz translation stage, was brought into close proximity with the end-face of the HC-PCF and aligned with the core center. A Navitar imaging system, mounted above the chip, allowed the insertion process to be monitored. In the experiment the insertion length was ~400 μm. The other end of the SMF was spliced to a wavelength division multiplexer (WDM) which combined light from a fiber-coupled 1064 nm laser with light from a custom-built fiber-coupled supercontinuum (SC) source (spectral width ~575 to 1000 nm, average total power tens of μW). The launched power was monitored in real-time using a photodiode or a power-meter. At the output end of the HC-PCF, the near-field profiles of the modes at different wavelengths were imaged using CCD cameras combined with optical filters. The transmitted optical power was monitored using a photodiode (PD2) and the spectrum measured by delivering the signal via a multi-mode fiber (MMF) to an optical spectrum analyzer (OSA). To ensure that mainly core-guided light was collected, a pinhole was used to block cladding light.

Fig. 1c shows the measured near-field intensity profile at the output of the HC-PCF, with ~3 mW of 1064 nm optical power launched into the tapered SMF and the nanospike placed at core center with an insertion length of ~400 μm. The intensity distribution almost perfectly matches that of an $LP_{01}$ mode, and the coupling efficiency was measured to be ~50%. The Fresnel reflection, determined by measuring the power at the 1% port of the 2×2 coupler just after the nanospike, was ~0.01% — more than an order of magnitude weaker than for a flat silica-water interface. At higher 1064 nm power the optomechanical trapping force starts to act, pulling the nanospike towards core center. To demonstrate this, the base of the nanospike (marked $\Delta$ in the inset of Fig. 2a) was initially offset a few μm from the core center and the optical trapping power increased to ~600 mW.

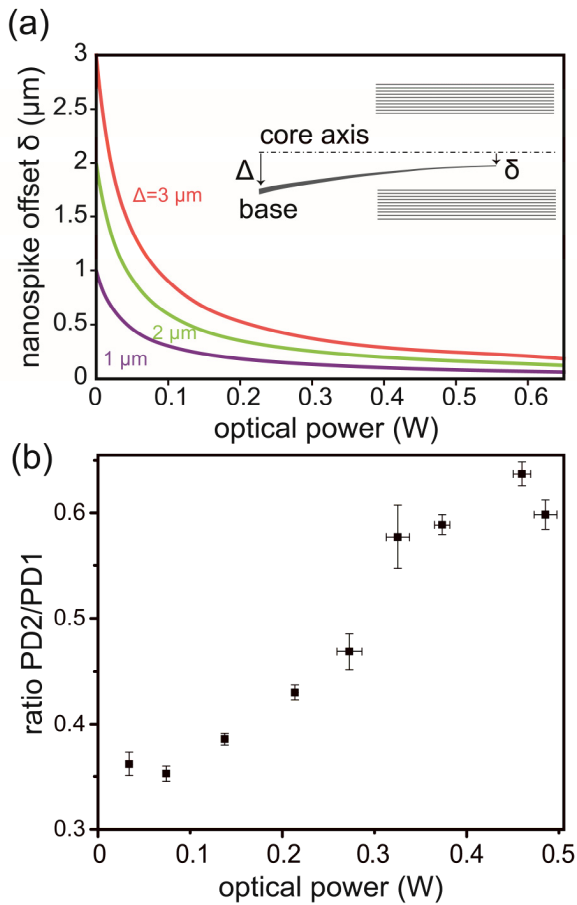

optical trapping force. Fig. 2a plots the calculated equilibrium position $\delta$, as a function of optical power, with $\Delta$ as parameter. For all values of $\Delta$, $\delta$ drops rapidly for increasing optical power, the effect eventually saturating at higher optical powers. This is because for higher optical powers the nanospike is pulled closer to the bottom of the optical trapping potential, where an increase in power causes only a small change in the equilibrium position. The saturation of equilibrium position with laser power is also a signature of an optical trapping force. Fig. 2b shows the experimentally measured relative transmission of the HC-PCF (ratio between the transmitted power at PD2 and the launched power at PD1) at 1064 nm for increasing trapping power. A smaller value of $\delta$ implies more efficient coupling to the low-loss fundamental mode of the HC-PCF and thus an increase in transmission.

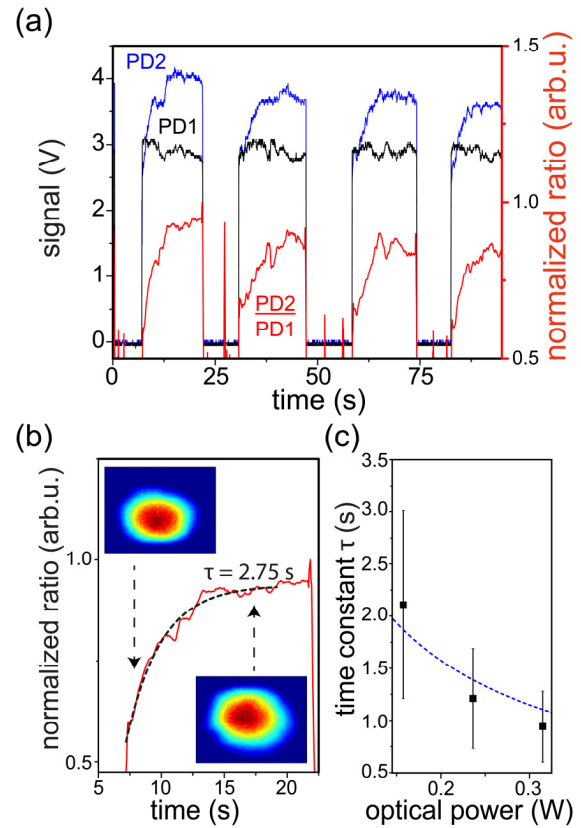

Fig. 3 (a) Signal collected at PD2 and PD1, and their ratio, while the 1064 nm laser was cycled on and off. (b) Zoom-in of the ratio PD2/PD1 from 5 to 25 s (red curve). An exponential fit (dashed black curve) yields a time-constant of 2.75 s. The average time constant over 11 consecutive measurements was measured to be τ = 2.81 ± 1.34 s. The inset shows the near-field mode profile at ~17.5 s. (c) The time-constant decreases as the optical power increases. The blue curve is a fit, with damping coefficient ϒ as a free parameter.

Fig. 2 (a) Simulated equilibrium position $\delta$ of the nanospike tip as a function of trapping power for several different values of base offset $\Delta$. (b) Ratio between the transmitted (PD2) and launched intensity (PD1) for increasing launched power.

The trapping force reduces the tip offset $\delta$ by an amount that is determined by the balance between the mechanical restoring force of the nanospike and the

The transmission increases significantly with optical power up to ~350 mW. For higher powers a clear trend towards saturation can be observed, agreeing well with the predictions of numerical simulations.

We further investigated the temporal response of the nanospike while cycling the 1064 nm trapping laser power on and off (Fig. 3a). When the trapping laser was switched on, the reference signal collected at PD1 showed an instant jump, remaining stable at a higher level thereafter (the fluctuations of the baseline are due to coupling instability into the SMF). In contrast, the transmitted signal (PD2) and the normalized relative transmission (PD2/PD1, red curve) increased over a timescale of several seconds, reaching a steady-state value. Repeated blocking and unblocking of the trapping laser beam confirmed that this behavior was reproducible and stable. Fig. 3b shows a zoom-in of the normalized relative transmission between 5 and 25 seconds. The signal increases by a factor of ~1.6, while the intensity profile of the core mode converges towards the $LP_{01}$ mode. Moreover, the increase in transmission can be fitted well to an exponential function with a time constant of $\tau = 2.75$ s.

The motion of the nanospike in response to an external force is strongly over-damped in liquid, i.e., the acceleration term in the equation of motion can be neglected. The resulting time constant can be written in the form $\tau = \Upsilon/(k_{mech}+k_{opt})$, where $\Upsilon$ is the viscous damping coefficient (N.s/m), $k_{mech}$ the mechanical stiffness (N/m), and $k_{opt}$ the optical spring stiffness (N/m), which is proportional to the trapping power. Fig. 3c plots $\tau$ versus trapping power for a nanospike similar to the one used in Fig. 3b. The three data-points are measurements, and as expected $\tau$ decreases as the optical power is ramped up. The mechanical stiffness $k_{mech}$ was calculated by finite-element modeling of the flexural eigenmode of the nanospike[22], yielding $k_{mech} \sim$ 1.4 pN/µm. The optical force (and thus the stiffness $k_{opt}$) can be calculated by integrating, along the nanospike, the gradient of the free energy of the local eigenmode with respect to the tip offset $\delta$[25], yielding a value of 20 pN/(µm.W). The fit to $\tau$ in Fig. 3c (dashed curve) was obtained for $\Upsilon \approx 8.5$ µN.s/m, in reasonable good agreement with the estimated value of 7.6 µN.s/m, calculated by modeling the nanospike as a row of interconnected spheres with varying diameters, calculating the Stokes drag on each and integrating along the nanospike[26,27]. The slightly higher value of $\Upsilon$ in the experiment may be explained by additional viscous drag caused by the proximity of the core walls[28].

Additionally to the 1064 nm trapping laser a broadband SC signal (~575 to 1000 nm) was launched into the nanospike using the WDM coupler. Fig. 4a shows the measured intensity profile of the core mode at wavelengths below 850 nm while the optical power of the trapping laser is ramped up. In this measurement the base of the nanospike was again offset 2 to 3 µm from core center. With increasing trapping power, the mode profile of the SC light converges towards the $LP_{01}$ mode. Also, in good agreement with the previous experiment and numerical simulations (Fig. 2), the mode profile saturates to a steady-state shape at higher trapping powers. Once the equilibrium position of the nanospike is reached, the excited mode profile remains stable over the course of the experiment.

When the nanospike is initially aligned to the core center, the optomechanical trapping force renders the launch efficiency robust against external perturbations. Fig. 4b compares the measured mode profiles when the base of the nanospike is displaced 2 µm from core center in the x- and y-directions (marked by red dashed squares), for low and high trapping power. At higher trapping power the mode profiles are less sensitive to base displacement. We attribute the asymmetry in mode profile versus displacement to a small initial offset of the nanospike from core center.

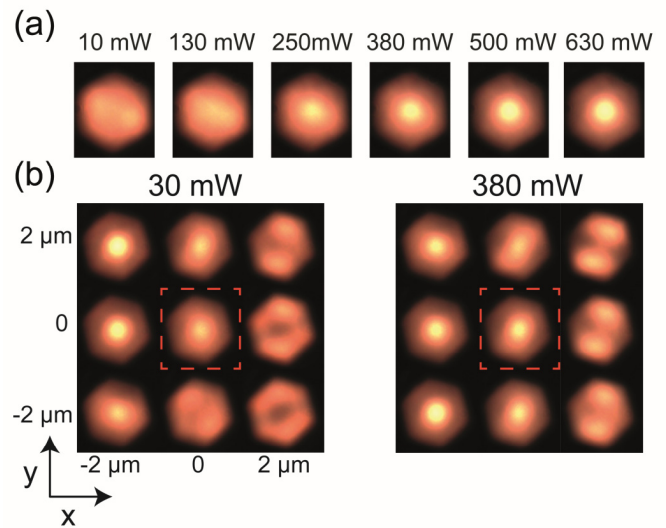

Fig. 4 Intensity profile of the SC for wavelengths below 850 nm, measured by a CCD camera for (a) a stepwise increasing trapping laser power, and (b) when the nanospike is initially trapped close to core center (indicated by red dashed box), and its base displaced transversally in steps of 2 µm.

The use of a SC source allows the wavelength dependence of the coupling efficiency to be characterized. Under optimal coupling conditions (i.e., with nanospike at core center), the SC coupling efficiency can be measured by dividing the spectrum directly after the WDM (black curve in Fig. 5a) by the transmitted spectrum after the HC-PCF (grey transparent curve), corrected for losses in the nanospike and the HC-PCF. To suppress contributions from intermodal-beating in the multi-mode fiber connection to the OSA, a smoothing filter was applied to the measured data (blue curve). The red curve in Fig. 5a plots the resulting coupling efficiency, showing that it remains almost achromatic from ~575 nm to 1000 nm, reaching ~40% for wavelengths longer than ~650 nm. The mode profile is $LP_{01}$-like at all wavelengths (Fig.

5b). These results agree quite well with numerical calculations (green dashed line) of the overlap integral between the $LP_{01}$-like mode of the HC-PCF and the mode profile at the end of the nanospike, assuming adiabatic evolution along the nanospike[29]. The spectral oscillations in the coupling efficiency (period ~50 nm) correspond to the estimated beat-period in the wavelength domain (~48 nm) between the core mode and the capillary mode of the first cladding layer of the HC-PCF[30].

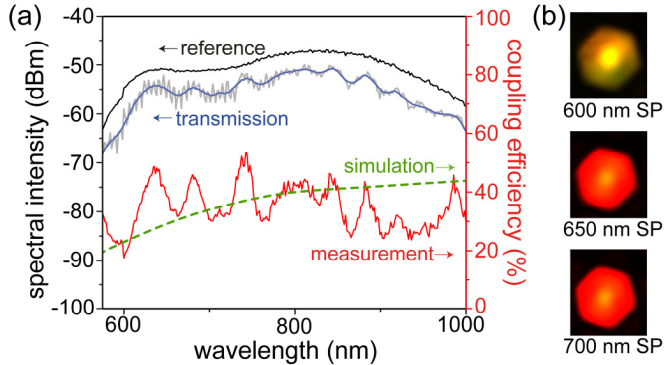

Fig. 5 (a) Wavelength dependence of the coupling efficiency (red curve), obtained by comparing the measured transmission spectrum after the HC-PCF (blue curve) with the known SC spectrum (black curve). The green dashed line plots the numerically calculated coupling efficiency (see text). (b) Near-field profiles of the core mode at several wavelengths, measured using the CCD camera and short-pass filters.

Finally, we tested the stability of the system against liquid counter-flow. The counter-flow was applied by raising the syringe pump ~36 cm above the launch end of the fiber, producing a flow rate of ~300 µm/s against the nanospike[31]. Fig. 6 compares the measured transmission signal (red curve, grey area) with the case of zero flow (black curve). The input signals are plotted as reference (upper curves). The launched optical power in this measurement was ~150 mW. It can be seen that the DC level of the transmission remains fairly constant after turning on the flow, indicating that the equilibrium position of the nanospike is negligibly perturbed by the counter-flow.

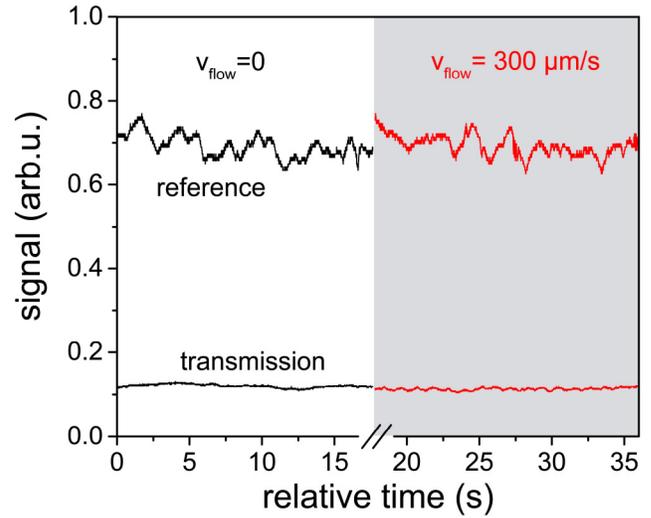

Fig. 6 Measured transmission signal for a counter-flow of ~300 µm/s (red curve) and zero flow (black curve). The corresponding laser input signals (measured at PD 1) are plotted as a reference (black and red upper curves).

## CONCLUDING REMARKS AND OUTLOOK

A fused silica nanospike, fabricated by simple thermal tapering and chemical etching, can be used for lensless, self-stabilized, broadband and efficient coupling to a liquid-filled HC-PCF. The measured Fresnel back-reflection is less than ~0.01 %, some 10 times less than at a flat glass-water interface and spectrally broader band than a typical anti-reflection coated lens. The coupling is also robust against a liquid counter-flow. While a kagomé-style HC-PCF was used to demonstrate the feasibility of the approach, the method can be potentially applied to all kinds of liquid-core or even gas-filled waveguides provided that the optical mode has a reasonable overlap with the mode of the nanospike and that the trapping force is sufficient to align the tip at moderate powers. Note that for smaller core diameters a substantially lower trapping power is required for self-alignment. The technique could be extended into the mid-infrared by suitable choice of soft-glass for the taper. We believe that the technique will allow integration of liquid-core waveguides into fiber-based microfluidic circuits, with applications in spectroscopy and photochemistry. The unique optomechanical system may also be exploited as a sensitive probe of hydrodynamic effects in liquid-core waveguides or for sensitive flow measurements in microfluidic systems[32].

## ACKNOWLEDGMENT

Richard Zeltner acknowledges funding from the Cluster of Excellence "Engineering of Advanced Materials" (www.eam.uni-erlangen.de) at the University of Erlangen- Nuremberg.